# A Data Warehouse Design for A Typical University Information System

Youssef Bassil

LACSC – Lebanese Association for Computational Sciences
Registered under No. 957, 2011, Beirut, Lebanon

**Abstract**

*Presently, large enterprises rely on database systems to manage their data and information. These databases are useful for conducting daily business transactions. However, the tight competition in the marketplace has led to the concept of data mining in which data are analyzed to derive effective business strategies and discover better ways in carrying out business. In order to perform data mining, regular databases must be converted into what so called informational databases also known as data warehouse. This paper presents a design model for building data warehouse for a typical university information system. It is based on transforming an operational database into an informational warehouse useful for decision makers to conduct data analysis, predication, and forecasting. The proposed model is based on four stages of data migration: Data extraction, data cleansing, data transforming, and data indexing and loading. The complete system is implemented under MS Access 2010 and is meant to serve as a repository of data for data mining operations.*

**Keywords**

*Data Warehouse, DBMS, Data Mining, Information System*

## 1. Introduction

Nowadays, almost every enterprise uses a database to store its vital data and information [1]. For instance, dynamic websites, accounting information systems, payroll systems, stock management systems all rely on internal databases as a container to store and manage their data. The competition in the marketplace has led business managers and directors to seek a new way to increase their profit and market power, and that by improving their decision making processes. In this sense, the idea of data warehouse and data mining was born [2]. In fact, data warehousing is the process of collecting data from operational functional databases, transforming, and then archiving them into special data repository called data warehouse with the goal of producing accurate and timely management information [3]; whereas, data mining is the process of discovering trends and patterns from data warehouse, useful to carry out data analysis [4].

A typical university often comprises a lot of subsystems crucial for its internal processes and operations. Examples of such subsystems include the student registration system, the payroll system, the accounting system, the course management system, the staff system, and many others. In essence, all these systems are connected to many underlying distributed databases that are employed for every day transactions and processes. However, universities rarely employ systems for handling data analysis, forecasting, prediction, and decision making. This paper proposes a data warehouse design for a typical university information system whose role is to help in and support decision making. The proposed design transforms the existing operational databases into an information database or data warehouse by cleaning and scrubbing the existing operational data. Besides, several columns and structures are dropped as they are useless for data mining. Finally, all captured and cleaned data are loaded and indexed in the warehouse making them ready for conducting data mining tasks.

## 2. Background

Operational Database: An operational database is a regular database meant to run the business on a current basis and support everyday transactions and processes [5].

Informational Database: An informational database is a special type of database that is designed to support decision making based on historical point-in-time and prediction data for complex queries and data mining applications. A data warehouse is an example of informational database.

Data Warehouse: It is a subject-oriented, integrated, time-variant, non-updatable collection of data used in support of management decision-making processes. It is subject-oriented as it studies a specific subject such as sales and customers behavior. It is integrated as it defines consistent naming conventions, formats, and encoding structures from multiple data sources. It is time-variant as it studies trends and changes over time. It is non-updatable as it is read-only, i.e. cannot be updated by regular users. [6]

Data Mining: In its broader sense, it is a knowledge discovery process that uses a blend of statistical, machine learning, and artificial intelligence techniques to detect trends and patterns from large data-sets, often represented as data warehouse. The purpose of data mining is to discover news facts about data helpful for decision makers [7].

## 3. Operational Database Design

Essentially, the operational database used to derive the data warehouse later on, encompasses fourteen distinct relations or tables associated together by means of relationships. It is a relational model database implemented under MS Access [8]. This database represents the business inside a typical university. It includes a front end registration system for handling students' registration processes, an accounting system for managing students' fees payments, a course management system for managing courses and assigning them particular sections and instructors, an alumni system for archiving students' records after graduation, and an assets system for distributing items such as machines, equipment, and computers over different departments. Figure 1 depicts the conceptual schema of the operational database.

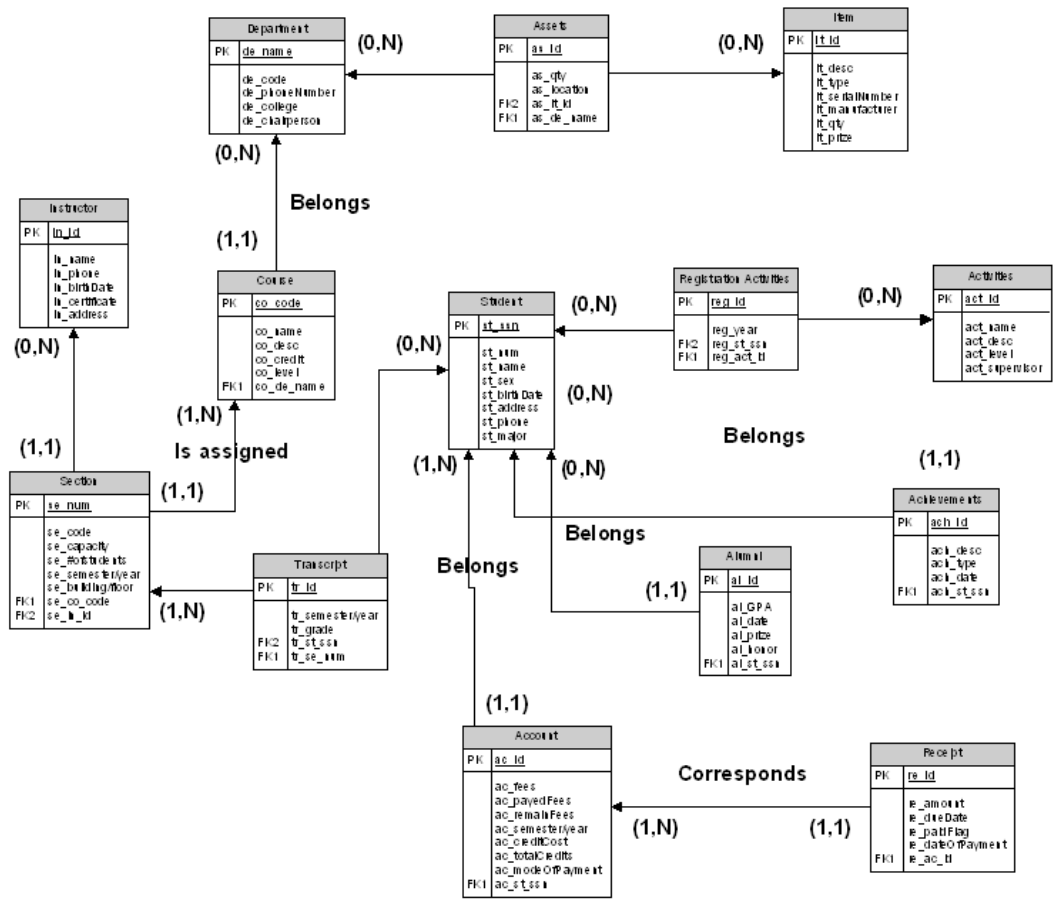

Figure 1 – Conceptual Schema of the Operation Database

## 4. Data Warehouse Design – Building the Informational Database

This section discusses the design of the proposed informational database along with its architecture, building process, SQL queries, and conceptual schema.

### 4.1. Architecture

In order to build the informational database, four vital steps are required to be completed, and they are respectively: Capture and extract, scrub and data cleansing, transform, and load and index. Figure 2 depicts the different stages required to transform the operational database into an informational database or data warehouse [9].

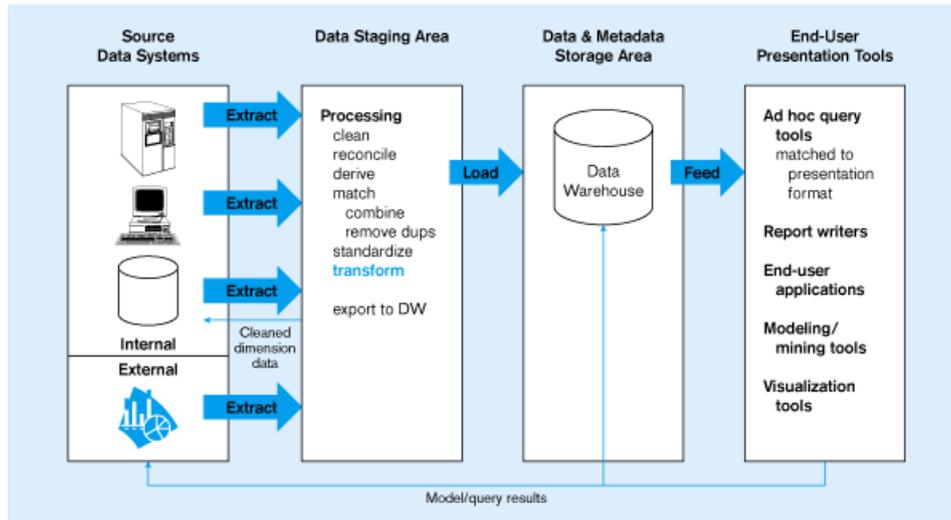

Figure 2 – Architecture for Building the Data Warehouse

Having the previously designed operational database as a data source, data are first extracted and then stored temporary into a buffer area. Once captured, data are pre-processed. This includes cleansing, scrubbing and reconciling data, fixing data entry errors, and transforming data into a more normalized standard. Once cleaned, the transformed data are loaded and indexed into the information database or the data warehouse. In this process, tables are dropped, new tables are created, columns are discarded, and new columns are added [10]. Figure 3 illustrates the building process of the data warehouse.

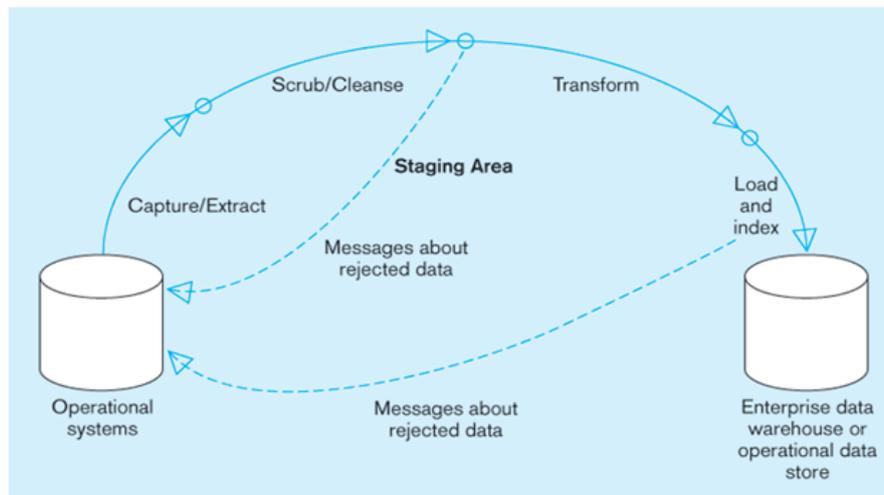

Figure 3 – Process of Building the Data Warehouse

### 4.2. Data Cleansing and Transforming

This section discusses all major transformations and SQL queries required in order to build and load the data warehouse.

### 4.2.1. Dropping Tables

Since decision-making is concerned with the trends related to students' history, behavior, and academic performance, tables "*assets*" and "*item*" are not needed; and therefore, they are discarded and excluded from the data warehouse.

*DROP TABLE assets ;*
*DROP TABLE item ;*

#### 4.2.2. Merging Tables

Based on the design assumptions, the three tables "*department*", "*section*", and "*course*" do not constitute separately important parameters for extracting relevant patterns and discovering knowledge. Therefore, they are merged altogether with the "*transcript_fact_table*" table.

SELECT co_name FROM course, section, transcript
WHERE tr_id = n AND str_semester/year = se_semester/year AND tr_se_num = se_num AND se_code = co_code ;

ALTER TABLE transcript fact table ADD co_course TEXT ;

DROP TABLE department ;
DROP TABLE section ;
DROP TABLE course ;

Furthermore, table "*Activities*" is merged with table "*RegistrationActivities*" and a new table is produced called "*RegisteredActivities*".

SELECT act_name FROM activities, registrationActivities
WHERE reg_act_id = act_id ;

#### 4.2.3. New Columns

During transformation new columns can be added. In fact, *tr_courseDifficulty* is added to table "*transcript_fact_table*" in order to increase the degree of knowledge and information.

ALTER TABLE transcript_fact_table ADD tr_courseDifficulty TEXT ;

Moreover a Boolean column is added to table "*receipt*" called *re_paidOnDueDate*

ALTER TABLE receipt (re_paidOnDueDate) ;

#### 4.2.4. Removing Columns

Unnecessary columns can be removed too during the transformation process. Below is a list of useless columns that were discarded during the transformation process from tables "*Account*", "*Student*", "*Receipt*" and "*Activities*" respectively:

ALTER TABLE Receipt REMOVE re_dueDate
                REMOVE re_dateOfPayment ;
ALTER TABLE Activities REMOVE ac_supervisor ;
ALTER TABLE Student REMOVE st_phone
                REMOVE st_email ;

### 4.3. Conceptual Schema – The Snowflake Schema

The proposed data warehouse is a Snowflake type design with one center fact table and seven dimensions [11]. Figure 4 reveals the basic Snowflake conceptual diagram of the proposed data warehouse.

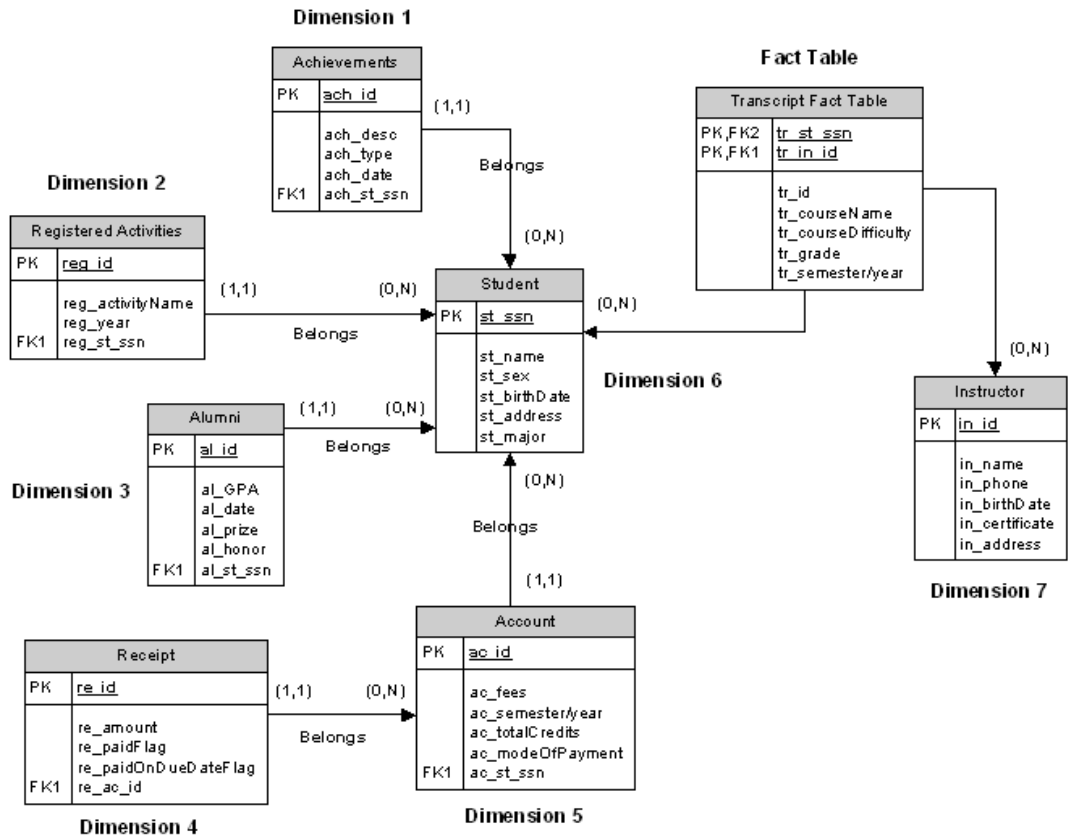

Figure 4 – Data Warehouse Snowflake Schema

## 5. Implementation: Loading and Indexing

The proposed data warehouse design is implemented under MS Access 2010. It contains eight refined distinct relations interrelated together. Figure 5 shows the logical diagram of the data warehouse implemented under MS Access. Now, as all data were successfully cleaned and transformed, they are exported from their buffer zone in which they were temporary stored to the new informational database or data warehouse just built. As a result, data mining can be carried out seamlessly in order to discover patterns and trends out of the warehouse that are necessary for business managers to perform decision making [12].

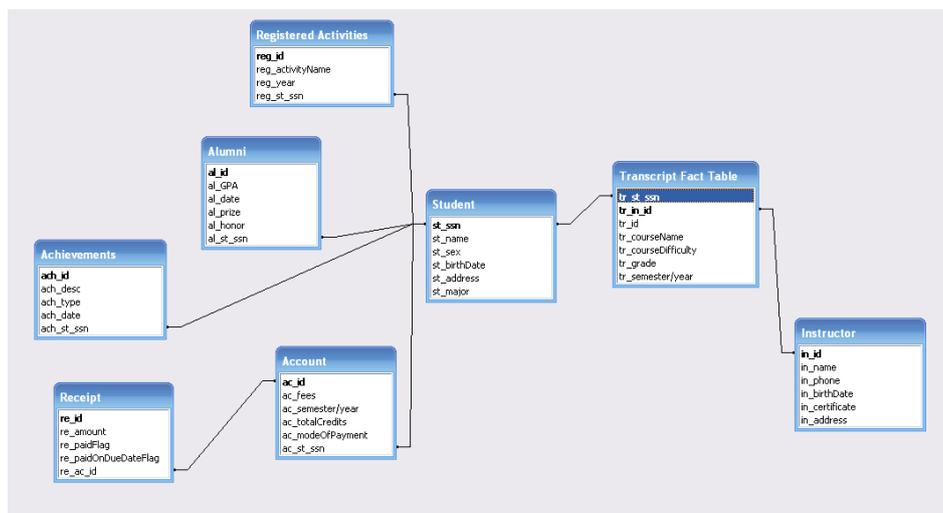

Figure 5 – Data Warehouse Logical Diagram under MS Access

16

## 6. Conclusions

This paper introduced a model for building a data warehouse for a typical university information system. The warehouse is an informational database whose data are extracted from an already existing operational database. The purpose of the proposed design is to help decision makers and university principles in performing data mining and data analysis over the data stored in the warehouse which eventually helps them in discovering critical patterns and trends.

## Acknowledgment

This research was funded by the Lebanese Association for Computational Sciences (LACSC), Beirut, Lebanon, under the "Data Warehouse Research Project – DWRP2012".